# The Compact Mirrors with High Pressure Plasmas

A.V. Anikeev [a], P.A. Bagryansky [a], A.A. Ivanov [a], A.A. Lizunov [a], S.V. Murakhtin [a],
V.V. Prikhodko [b], S. Collatz [c] and K. Noack [c]

[a] *Budker Institute of Nuclear Physics, Prospect Lavrent'eva 11, 630090 Novosibirsk, Russia.*
[b] *Novosibirsk State Universyty, Pirogova str. 2, 630090 Novosibirsk, Russia.*
[c] *Forschungszentrum Rossendorf, Postfach 51 01 19, 01314 Dresden, Germany.*

## 1. Introduction

The Gas Dynamic Trap (GDT) experimental facility at the Budker Institute Novosibirsk is a long axial-symmetric mirror system with a high mirror ratio variable in the range of 12.5-100 for the confinement of a two-component plasma [1]. One component is a collisional "background" (or "target") plasma with ion and electron temperatures up to 100 eV and density up to $10^{14}$ cm$^{-3}$. The ion mean free path of scattering into the loss cone is much less, than the mirror-to-mirror distance for this component that results in the gas-dynamic regime of confinement. The second component is the population of high-energetic, so-called fast ions with energies of 2-18 keV and a density up to $10^{13}$ cm$^{-3}$, which is produced by 45° neutral beams injection (NBI) in the middle of the central cell [2]. The fast ions are confined in the mirror regime having their turning points at the mirror ratio of 2. To provide MHD stability of the entire plasma axial-symmetric min-B cells are attached to both ends of the device.

At present, the GDT facility is being upgraded. The first stage of the upgrade is the **S**ynthesized **H**ot **I**on **P**lasmoid (**SHIP**) experiment. It aims, on the one hand, at the investigation of plasmas, which are expected to appear in the region of high neutron production in a GDT based fusion neutron source proposed by the Budker Institute [3] and, on the other hand, at the investigation of plasmas the parameters of which have never been achieved before in magnetic mirrors. The expected record values of plasma parameters and several peculiarities of the plasma, like the composition of two energetically very different ion components where the high-energetic part represents the majority, strong non-isotropic angular distribution of the high-energetic ions and non-linear effects as non-paraxial effective magnetic field and high value of β offer a great field for interesting investigations.

The construction of SHIP is now completed and the first experimental activities were already started in this year. The paper presents the physical concept of the SHIP experiment, the results of numerical pre-calculations and draws conclusions regarding possible scenarios of experiments.

## 2. Technical description and Scientific Objectives

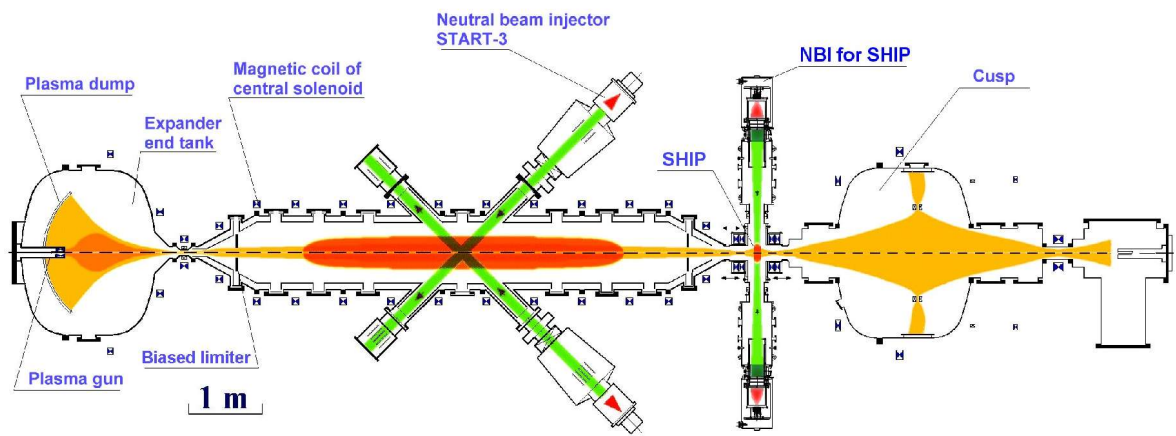

***Fig. 1***: *GDT facility with the SHIP experiment.*

The experiments will be performed in a small, additional mirror section, which is installed at the end of one side of the GDT. Fig. 1 shows the schematic layout of the GDT-SHIP experiment. The magnetic field on axis will be in the range of 0.5-5.2 Tesla and the mirror ratio will amount to 1.2-1.4. The magnetic field strength will be varied by extending/shortening the distance between the SHIP mirror coils. The SHIP volume is filled with background plasma with a density of about $5 \times 10^{13}$ cm$^{-3}$ streaming in from the central cell of the GDT. This plasma component is maxwellized and has an electron temperature of about 100 eV. It is pre-heated up by the standard NBI system of the GDT. Two newly developed neutral beam injectors will perpendicularly inject into the SHIP mirror a total current up to 120 eq. Amperes of hydrogen or deuterium atoms with an energy up to 25 keV as pulse with a duration up to 3 ms. Ionization of the beams by the target plasma generates the high-energetic ion component. The density of the resulting Hot Ion Plasmoid is expected to be considerably higher than that of the target plasma. For the given experimental conditions, the lifetime of the synthesized plasma is essentially determined by the target plasma cooling rate and might be of the order of one millisecond. Since the energy loss of the fast ions by background plasma with high electron temperature is negligible in the millisecond time-scale the averaged energy of the trapped ions is expected to be not much lower than the injection energy, i.e. in the range 15-20 keV. It was estimated that their density will reach $10^{14}$ cm$^{-3}$ in a volume of about 500 cm$^3$ even in the case of low magnetic field what will result in high β-values between 20-60 per cent.

Except these record values of plasma parameters also several peculiarities of the plasma, like the composition of two energetically very different ion components where the high-energetic part represents the majority, strong non-isotropic angular distribution of the high-energetic ions and non-linear effects as non-paraxial effective magnetic field and high value of β offer a great field for interesting investigations. The high plasma parameters and the stated peculiarities raise several fundamental questions:

- What are the attainable maximum parameters of stable background plasma and of fast ions?
- Does appear a high-β threshold to instability of any kind in the attainable parameter range?
- Do appear self-organizing effects in such plasmas and what are the consequences regarding equilibrium distributions and stability?
- To which degree does the non-paraxial magnetic field influence on equilibrium and stability of both plasma components?

The answers to these questions are the objectives of the experimental and theoretical research of the SHIP project.

## 3. Numerical Simulations of the SHIP.

The numerical methods and codes that are to be applied for SHIP simulations concern the Integrated Transport Code System (ITCS) which has been developed in recent years in collaboration between the Institute of Safety Research of the Forschungszentrum Rossendorf and the Budker Institute [4]. At present, this system allows the interactive calculations of target plasma, neutral gas and of fast ions in the central cell of the GDT facility in the frame of classical plasma theory. In addition, the ITCS can calculate the resulting neutron production by D-D or D-T fusion reactions in case of deuterium/tritium neutral beam injection.

The linear version of the **M**onte **C**arlo **F**ast **I**on **T**ransport code MCFIT, that is the main part of ITCS, simulates the transport of neutral beam produced high-energetic ions interacting with a given magnetic field, a target plasma and with neutral gas [5]. The code stochastically generates independent ion histories, which during their lifetimes contribute to the estimations of the quantities of interest. The main disadvantage of the method is the slow convergence of the statistical error according to $N^{-1/2}$ where $N$ is the number of simulated particle histories. On the other side, the code describes the relevant transport processes with a minimum of approximations. The assumption of a linear fast ion transport turned out to be a good approximation for the experiments that were performed up to now at the GDT.

As it was already pointed out, in SHIP experiments the plasma physical situation will be substantially different from that in the GDT. Here, the fast ion density is expected to be remarkably higher than that of the target plasma ions. In contrast to this relation, in GDT experiments the target plasma ion density was about one order of magnitude higher than the fast ion density. This fact has the consequence that several interactions of the fast ions become non-linear that means that the "interaction partners" - i.e. magnetic field, background plasma and neutral gas – now depend on the fast ion field too. To meet such requirements in the simulation a Monte Carlo code offers only the possibility to do this by means of iterations: The first simulation is done with pre-defined start values of the "partners" and gives the first approximation of the fast ion field. After calculating new values of the "partners" the next fast ion simulation follows. To prepare the MCFIT code for that purpose several modifications had to be introduced:

### 3.1 Splitting of the ion density

A strong fast ion density produces a high electrostatic potential that decreases the ion density of the target plasma. In an approximate model the following relationship may be derived

$$n_w = (\sqrt{n_f^2 + 4 \cdot n_0^2} - n_f)/2$$
$$n_e = n_w + n_f$$

Here are $n_w$ - density of the so called "warm" ions of the target plasma, $n_f$ - density of the fast ions, $n_e$ - electron density, $n_0$ - unperturbed plasma density. All densities are in their radial ($r$), axial ($z$) and time ($t$) dependencies and $n_0(r,t)$ is a fixed radial profile of the target plasma at the entrance into the SHIP chamber. The assumptions of the model are: the fixed profile $n_0$ for electrons and warm ions at the entrance, their distribution according to the law of Boltzmann with the same temperatures $T_w$ and $T_e$ in the region of sloshing fast ions and the neutrality condition. The newly introduced fast and ion density $n_f$ gives now a contribution to the ionization process of the neutral beams and to the angular scattering of the fast ions.

Figures 2 a) and b) illustrate the steady state radial (in the mid-plane) and axial (on the axis) distributions of the density of different plasma components in SHIP as they are to be simulated for the variant *v.1* in Table 1. One can see that the warm ion density is essentially reduced by the fast ion potential even for this scenario with not very high parameters.

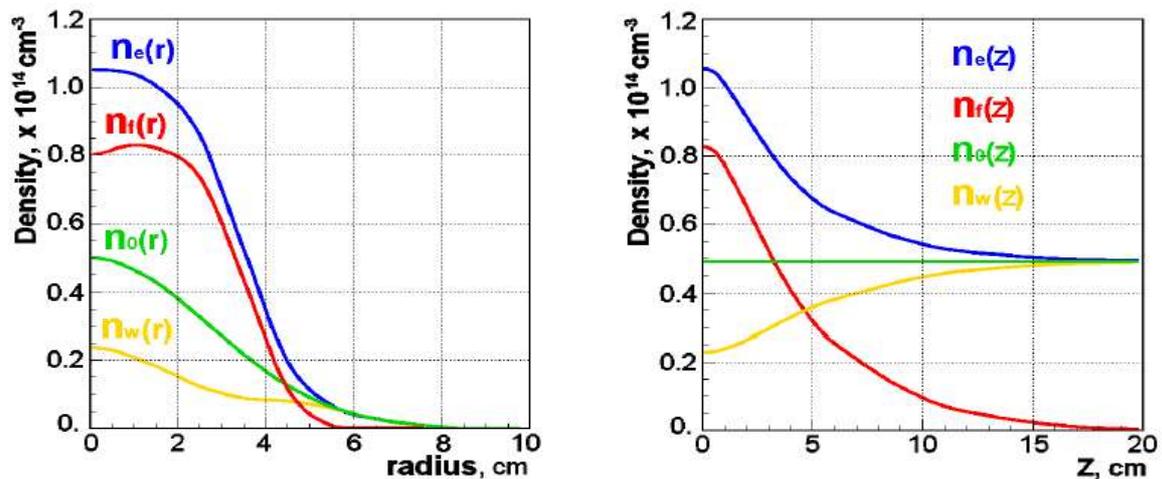

***Fig. 2*** *a) Radial distributions of densities.      b) Axial distributions of densities.*

*3.2 Calculation of high β effects in SHIP*

In the SHIP experiment and also in the GDT-Upgrade the high fast ion energy content results in a high value of β which reaches almost sixty per cent. The high-β effect causes a deformation of the vacuum magnetic field and, consequently, of the fast ion distribution too. To consider this non-linear effect the time and spatial distribution of azimuthal fast ion currents, calculated by MCFIT, were used to compute the correction of the magnetic field according to the Biot-Savart law. Then, this β-corrected, time dependent magnetic field was used by MCFIT in an iteration procedure.

For example, the comparison of high-β-corrupted and vacuum magnetic fields in the SHIP is presented on Fig.3. The β-corruption of fast ion density profile is shown on Fig.4.

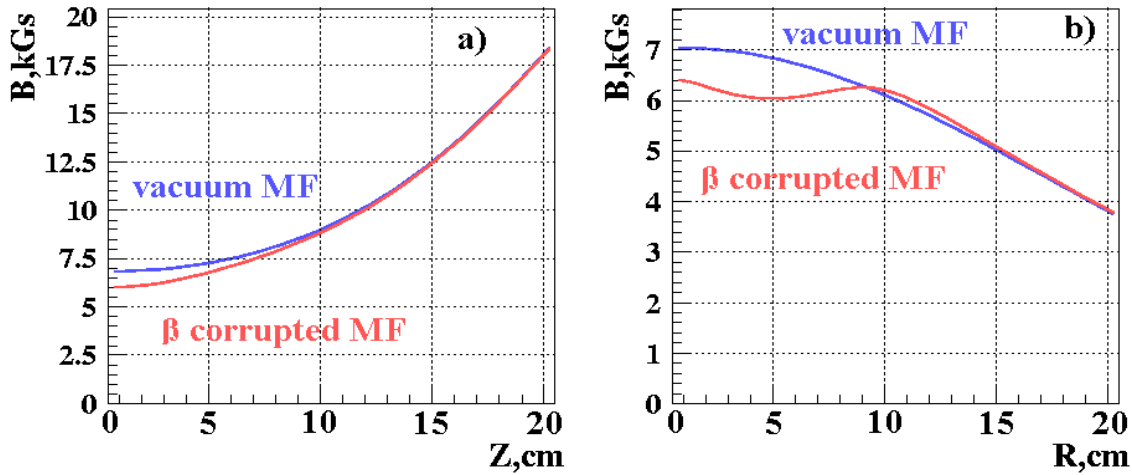

***Fig.3*** *The β-corrupted magnetic field: a) on-axis profile; b) radial profile. The blue curves are correspond to vacuum magnetic field. Z=0 is the midplane of the SHIP device.*

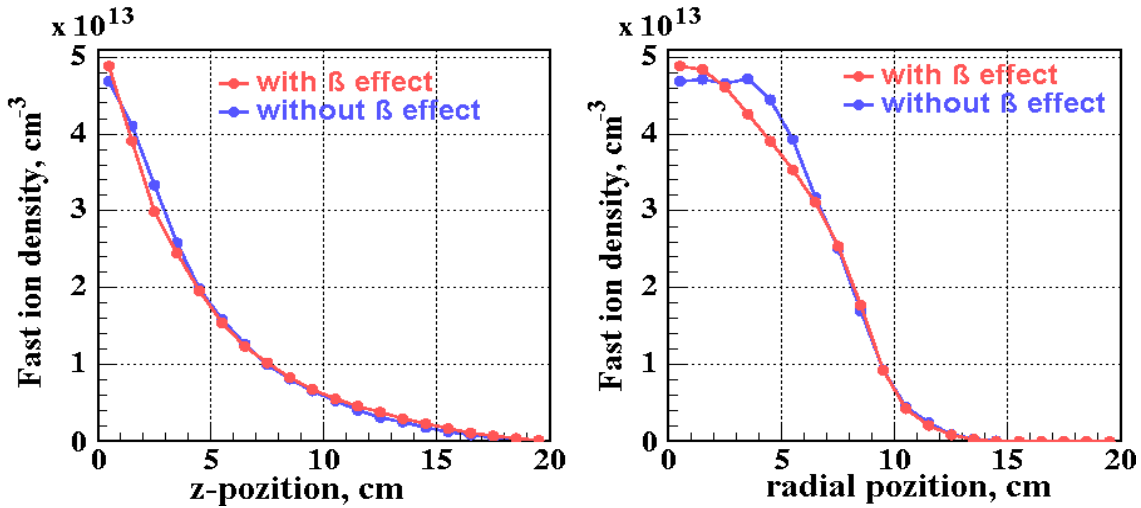

***Fig.4***. *Influence of the high-β on fast ion density profile in the SHIP.*

*3.3 Calculation of the neutral gas distribution in SHIP.*

For the calculation of the density distribution of fast (high-energetic) ions in a gas dynamic trap it is necessary to take into account their interaction with the neutral gas inside the vacuum chamber. The computer code NEUSI [6] calculates the distribution of the neutral gas components in large devices like the GDT. It makes use of the approximative assumption that the radius of the chamber is small in comparison with its length. This approximation is not justified for the SHIP device. Therefore, the new neutral gas code NEUFIT has been developed which avoids this approximation.

In SHIP the following sources of neutral gas components appear (see Fig 5). The neutral hydrogen atoms injected as neutral beams interact with the ions of the thermal plasma. The charge exchange process produces the fast ions and slow neutral hydrogen atoms which represent the primary gas source. During their life histories the fast ions interact with neutral gas components and by charge exchanging with slow neutral atoms they will become fast neutral atoms. Also, the thermal plasma interacts with neutral gas components and produces fast and slow neutrals.

A part of the neutral atoms reaches the wall of the vacuum chamber. Some of them will be reflected with reduced energy. NEUFIT approximately assumes that all reflected atoms are slow ones. The other part accumulates on the wall surface as hydrogen molecules and returns with low energy into the chamber volume. In this way, we have tree components of neutral gas in the facility: slow hydrogen atoms, fast hydrogen atoms and slow hydrogen molecules. NEUFIT calculates the density distributions of these three gas components over the time of an experimental shot with the help of a computation procedure for given time courses of fast and slow ion distributions.

In the Fig.5 the calculated density distributions of the neutral gas components in the midplane of the SHIP device are shown.

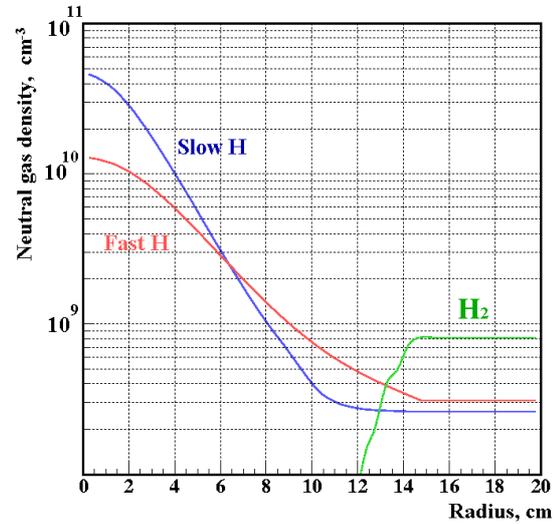

**Fig.5** Neutral gas density distributions in the SHIP midplane.

### 3.4 Results.

Up to now four regimes with different input parameters were considered. The assumed parameters of the experiments and the results of calculations are presented in Table 1.

*Table 1*: *Parameters of proposed SHIP experiments and results of calculations.*

| **Parameters** | v.1 | v.2 | v.3 | v.4 |
|---|---|---|---|---|
| Length/cm | 43 | 43 | 43 | 58 |
| Magnetic field/T: midplane/mirror | 2.3/5.2 | 2.3/5.2 | 2.3/5.2 | 0.7/2.5 |
| *Target plasma:* | | | | |
| electron temperature $T_e$/eV | 80 | 100 | 80 | 100 |
| unperturbed density $n_0$ / $10^{14}$cm$^{-3}$ | 0.5 | 0.5 | 0.5 | 0.5 |
| warm ion density $n_w$ / $10^{14}$cm$^{-3}$ | 0.2 | 0.18 | 0.12 | 0.3 |
| electron density $n_e$ / $10^{14}$cm$^{-3}$ | 1.1 | 1.38 | **2.02** | 0.8 |
| *NBI System:* | H | H | D | H |
| energy/keV | 25 | 25 | 25 | 25 |
| total equivalent current / A | 80 | 120 | 80 | 80 |
| injected power / MW | 2 | 3 | 2 | 2 |
| duration / ms | 1.5 | 2 | 1.5 | 2 |
| *Fast Ions:* | | | | |
| maximal density $n_f$/ $10^{14}$cm$^{-3}$/ | 0.85 | 1.2 | **1.9** | 0.5 |
| mean energy / keV | 10 | 14 | 9 | 12 |
| total energy contents / J | 81 | 132 | **166** | 120 |
| trapped power / kW | 320 | 680 | 675 | 650 |
| drag power / kW | 295 | 480 | 640 | 520 |
| cx loss power/kW | 25 | 200 | 35 | 106 |
| maximal β/% | 8 | 12 | 15 | **62** |

The interest was focused on the selection of experimental scenarios with the maximal fast ion density (*Variant 2 and 3 in Table*), comparison of D and H injection (*Variant 1 and 3*) and maximal local β parameter (*Variant 4*) for numerical study of high-β effects in SHIP. The variant *3* correspond to most preferable parameters of the SHIP experiment with maximum of fast ion density and energy contents. Moreover, the injection of deuterium allows to measure reaction products that gives padding diagnostic capabilities. The total rate of DD reactions in the SHIP experiment was obtained as about $8\times10^{11}$ neutron/s in this numerical simulation. The time dependences of the main SHIP parameters for the variant with D neutral injection are shown on Fig.6. It turned out, that the fast ion population reached its steady states very quickly, already in about 1 ms.

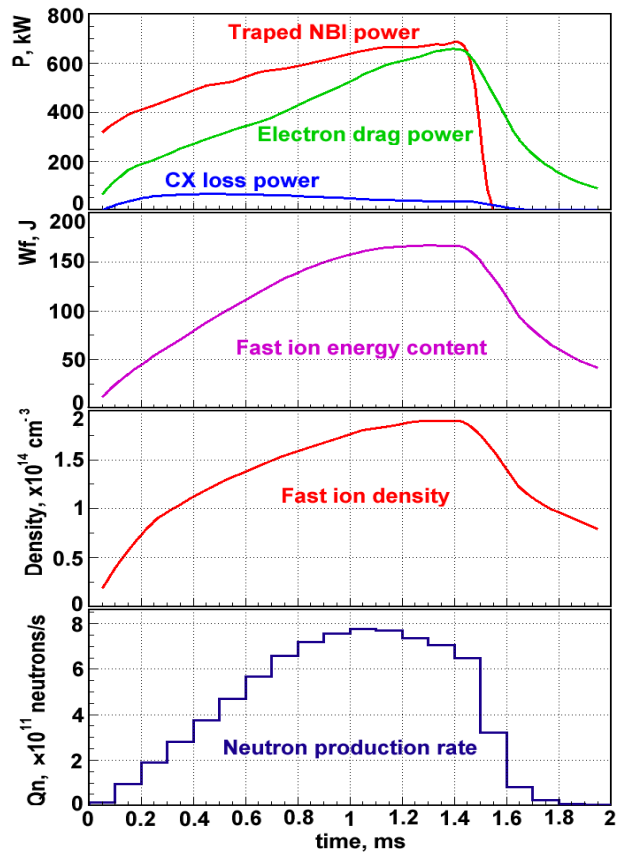

**Fig. 6** *Time evolution of the main parameters for the variant 4 with D neutral injection.*

**4. First SHIP experimental activity.**

The construction of SHIP is now completed and the first experiments were started at GDT-SHIP device. The two 20 keV $D^0$ beams with total current of 20 eq. A were injected to the SHIP mirror cell with warm target plasma ($T_e\sim50$ eV, $n_e\sim10^{13}$ cm$^{-3}$). The basic diagnostics and operating system were tested in this experimental session.

Figure 7 presents the linear density of target plasma in the SHIP cell as a 8 mm interferometer data. The signal of diamagnetic loop in SHIP is shown on Fig.8. Maximal value of plasma diamagnetism corresponds to the fast ions total energy content of about 4 J. Low parameters of the first SHIP experiment is effect of high charge-exchange losses of fast ions and moderate SHIP-NBI power.

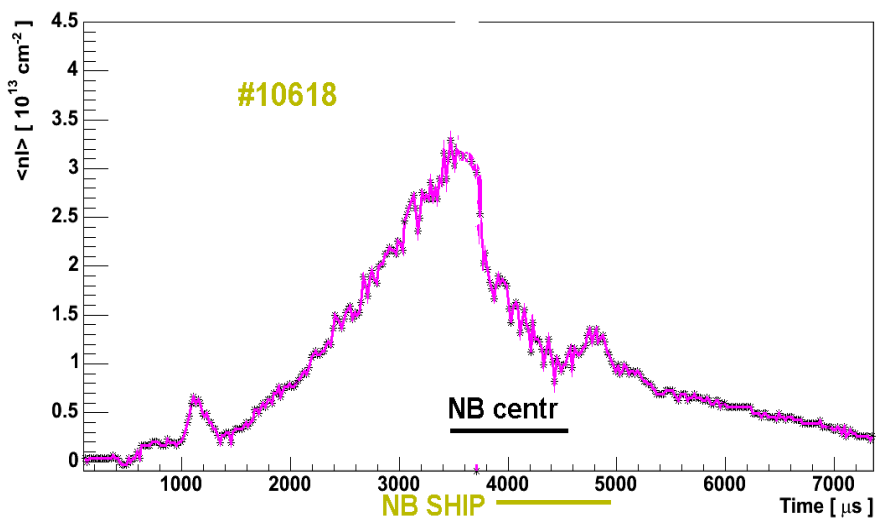

***Fig.5*** *The linear density of target plasma in SHIP. (8mm interferometr data)*

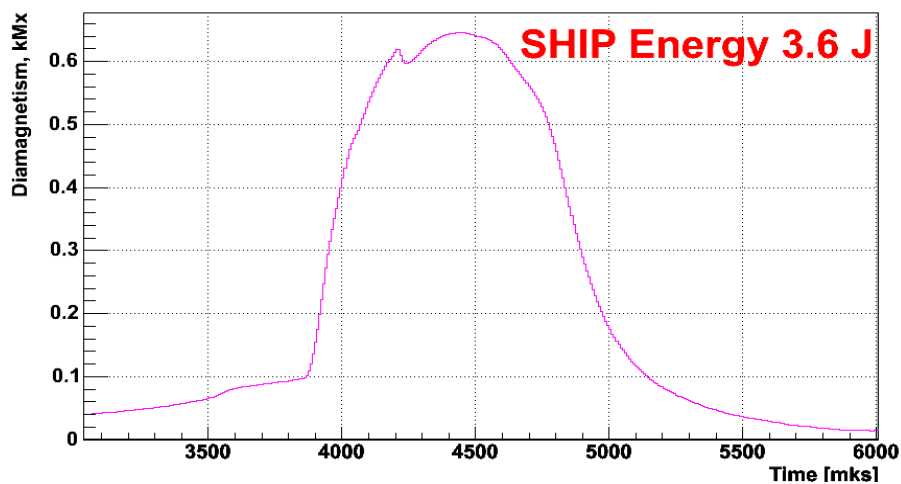
*Fig. 6* *The SHIP diamagnetism. (Loop data)*

The next step of experimental activity is installation of the new SHIP NBI power supply system (2MW for two 25 keV beams, 2 ms pulse duration) and reduction of the CX losses by improvement of the first wall condition. It allows obtaining the high plasma parameters in SHIP that were simulated in presented work.

**5. Conclusions.**

From the work presented in this paper the following conclusions may be drawn:
- The SHIP experimental activity has been started in this year.
- The fast ion transport code MCFIT has been extended to be able to consider non-linear processes that are expected to be of importance in SHIP experiments. Than, it was used to study certain possible experimental scenarios.
- The new code NEUFIT was developed for gas simulation inside the SHIP chamber and was used for numerical simulations together with MCFIT.
- The simulation of a maximal NBI power regime with hydrogen injection gave the fast ion density $1.2 \times 10^{14}$ cm$^{-3}$ with a mean energy of 14 keV.
- The calculation of the deuterium injection regime with 2 MW NBI power gave the maximal fast ion density of $1.9 \times 10^{14}$ cm$^{-3}$ with the mean energy of 9 keV.
- The calculation of an experimental scenario with reduced magnetic field resulted in a maximal β-value of 62%. So, this regime is recommended for the study of high-β effects in plasmas confined in axial-symmetric mirrors.

**Acknowledgments**

This work was in part supported by the STC agreement between Germany and Russia through the project RUS 01/580. Presented author A.V. Anikeev is grateful to INTAS and to the Alexander von Humboldt Foundation for supporting his stays at Forschungszentrum Rossendorf, Germany.